\documentclass[journal,10pt]{IEEEtran}
\usepackage[T1]{fontenc}
\usepackage[utf8]{inputenc}
\usepackage{amsmath,amssymb,graphicx,booktabs,array,tabularx,url,cite}
\usepackage{microtype}
\usepackage{balance}
\graphicspath{{figures/}}

\begin{document}
\title{Scenario-Based Assessment of Transport Electrification for Low-Carbon Urban Planning: A Case Study of Phu Quoc, Vietnam}
\author{Tien-Khai Nguyen\thanks{The case study was originally developed in 2022 within an urban-infrastructure planning project at Ho Chi Minh City University of Architecture, Vietnam. Corresponding author: Tien-Khai Nguyen (e-mail: khainguyentien555@gmail.com). This revised manuscript distinguishes source-reported results from independent consistency checks.}}
\maketitle
\begin{abstract}
Urban net-zero strategies require a defensible emissions boundary and quantitative estimates of the contribution achievable through individual mitigation measures. This study assesses road-transport electrification within a planned residential development in Phu Quoc, Vietnam, using the assumptions and scenario data of an urban-planning study originally prepared in 2022. A vehicle-class inventory combines population-dependent fleet projections, assumed travel activity, gasoline and diesel consumption, fuel-specific carbon dioxide (CO$_2$) factors, and electricity-related emissions from battery-electric vehicles. The reference case projects annual road-transport emissions of 59.41, 73.86, and 87.23 tCO$_2$ for 2022, 2025, and 2030, respectively. Three year-specific substitution scenarios, corresponding to nominal electric-vehicle shares of 25\%, 50\%, and 100\%, yield source-reported avoided emissions of 9.43, 13.64, and 55.41 tCO$_2$/year, or 15.9\%, 18.5\%, and 63.5\% of their respective baselines. However, a cross-scenario audit identifies an inconsistency in the 2025 mixed-fleet results under time-invariant vehicle-class emission-intensity ratios: under that additional assumption, 2025 avoided emissions would be 23.45 rather than 13.64 tCO$_2$/year. The original detailed vehicle-energy input tables are not included in the available source document, preventing definitive reconciliation. The study therefore provides a transparent, preliminary planning assessment rather than evidence of citywide net-zero attainment. Its principal value is a reproducible accounting structure and an explicit identification of the data, electricity-system, and infrastructure constraints that must be resolved before policy or investment decisions.
\end{abstract}
\begin{IEEEkeywords}
Battery electric vehicles, carbon accounting, electric mobility, emission inventory, low-carbon cities, net-zero planning, Phu Quoc, scenario analysis, urban transportation.
\end{IEEEkeywords}

\section{Introduction}
\IEEEPARstart{U}{rban} development increases demand for transport, buildings, materials, water, and municipal services. These activities generate direct emissions within a development as well as indirect emissions associated with imported energy and supply chains. The notion of a ``net-zero city'' is therefore more demanding than the elimination of vehicle exhaust: it requires an explicitly defined geographic and accounting boundary, comprehensive source coverage, quantified reductions, and the treatment of residual emissions and removals. The Intergovernmental Panel on Climate Change (IPCC) defines net-zero greenhouse-gas emissions in terms of a balance between anthropogenic emissions and removals over a specified period \cite{ipccgloss}; community inventories likewise distinguish emissions released locally from those attributable to purchased energy and activities beyond the urban boundary \cite{gpc}.

The transition from internal-combustion-engine (ICE) vehicles to battery-electric vehicles (BEVs) is a relevant intervention because it eliminates tailpipe emissions while transferring part of the emissions burden to electricity generation. The benefit is consequently conditional on vehicle energy use, travel demand, and the grid's emissions intensity. The International Energy Agency (IEA) identifies road-transport electrification and electricity-system decarbonization as complementary elements of the mobility transition \cite{iea2026}. This distinction is particularly important for island cities, where tourism and infrastructure development can change both transport activity and power-system demand.

Phu Quoc, a Vietnamese island-city tourism center, provides a useful planning setting. Long-term planning instruments have emphasized sustainable urban development, tourism, conservation, and supporting infrastructure \cite{decision633,decision486}. Within that setting, the original 2022 study considered a residential development in Duong To and assessed the emissions consequences of progressive electric mobility \cite{originalstudy}. The present work converts that material into a clearly bounded, transport-focused scientific assessment. It does \emph{not} claim to demonstrate a complete urban net-zero solution.

The contributions are threefold. First, a city-level conceptual framework is linked to an operationally specified road-transport CO$_2$ calculation rather than to a purely aspirational target. Second, the source fleet, mobility, fuel, and electrification assumptions are consolidated into a consistent technical presentation with explicit units and boundaries. Third, a numerical audit tests whether the reported scenarios can be reconciled with a common vehicle-class energy-intensity model, thereby separating planning insight from unresolved input-data limitations.

\section{Urban Net-Zero Framework and Assessment Scope}
\subsection{Definition and Accounting Boundary}
Net-zero emissions should be differentiated from ``zero emissions at the point of use.'' A BEV produces no CO$_2$ exhaust during driving, but the electricity used for charging may be associated with emissions at generation facilities. Further emissions arise from vehicle and battery manufacture, road construction, building operation, tourism, solid waste, and end-of-life activities. An appropriate urban inventory should therefore state which activities are counted and how imported electricity is treated \cite{ipccgloss,gpc}.

\begin{figure}[!t]
\centering\includegraphics[width=\columnwidth]{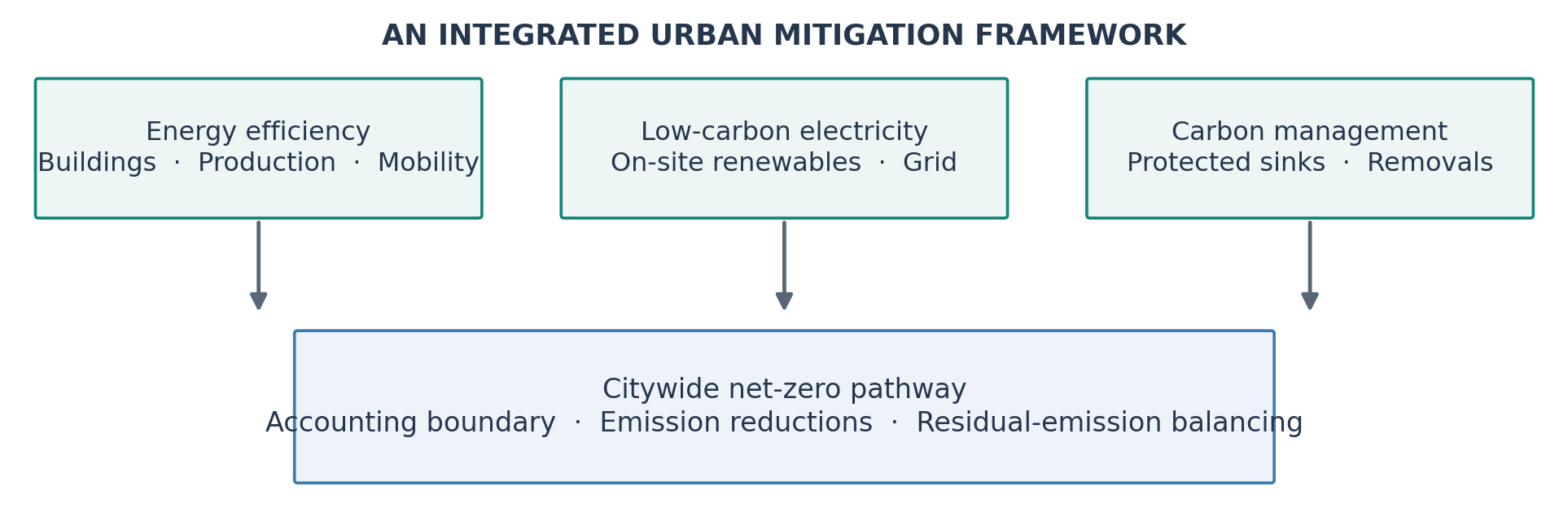}
\caption{Conceptual net-zero urban framework redrawn in English from the source study. The framework is a planning taxonomy; no citywide balance is quantified here.}\label{fig:netzero}
\end{figure}

Figure~\ref{fig:netzero} synthesizes the three broad pillars discussed in the original project: efficient urban energy use; low-carbon electricity and renewable energy; and carbon management, including durable carbon sinks and treatment of residual emissions. Efficiency interventions span production, construction, buildings, and mobility. Green-space conservation may support ecosystem protection and carbon storage, but it does not automatically offset quantified urban emissions. Similarly, green-building certification is an assessment framework, not proof that embodied or operational emissions are zero.

\subsection{Emission Sources and Complementary Measures}
\begin{figure}[!t]
\centering\includegraphics[width=\columnwidth]{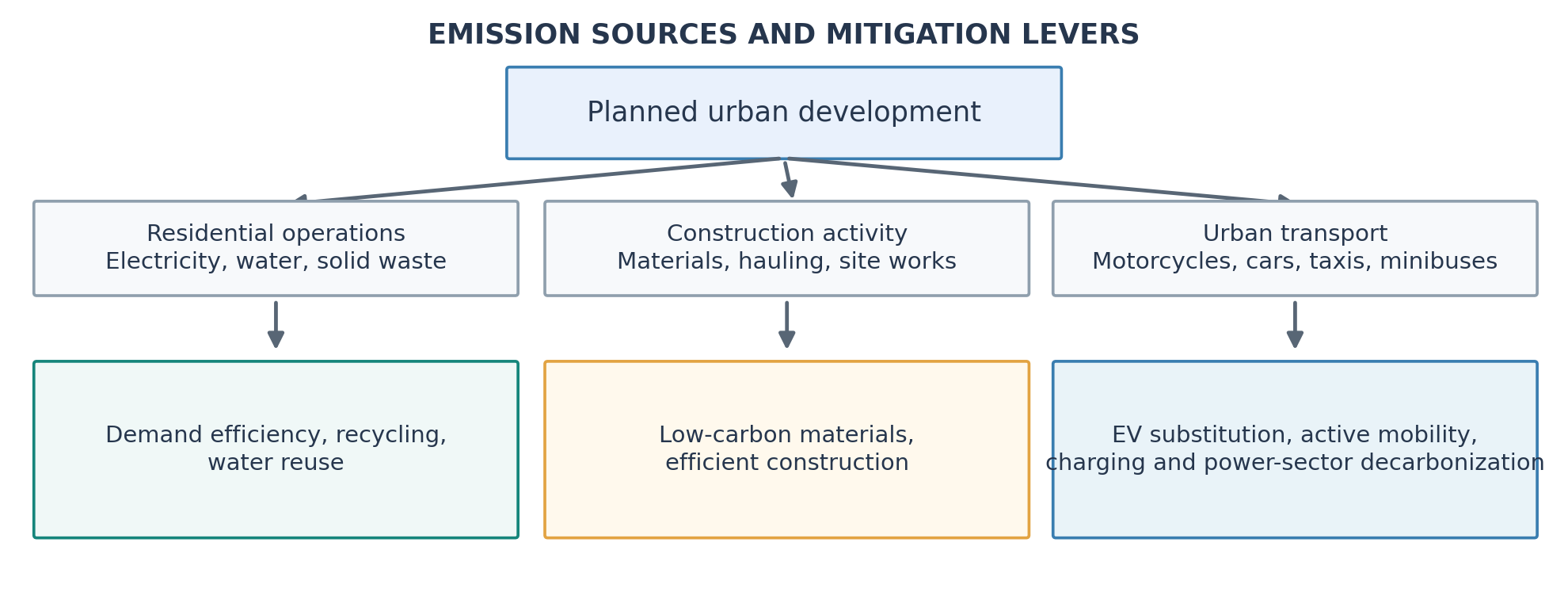}
\caption{Development emissions and possible intervention pathways. Only the road-transport CO$_2$ pathway is quantified in this study.}\label{fig:source}
\end{figure}

The planned area contains 185 terraced houses accommodating an initial projected population of 925 residents \cite{originalstudy}. The source study identified three principal emission domains: (i) residential operation, including electricity, water supply and wastewater, and municipal solid waste; (ii) construction, including building materials, haulage, dust-related site activity, and construction debris; and (iii) road transport associated with motorcycles, cars, taxis, and minibuses. Tourism adds demand for mobility, accommodation, and associated services. Figure~\ref{fig:source} organizes these sources and the corresponding interventions in English.

For residential activity, proposed interventions include lower energy demand, water reuse, separation and recycling of waste, and efficient utilities. For construction, the source recommends recycled or lower-impact materials, material-efficient design, and green-building assessment systems such as LOTUS, LEED, Green Mark, and EDGE. For travel, it considers cycling and pedestrian access, electric mobility, and lower-carbon fuels. For tourism, it discusses environmental monitoring, conservation, waste and wastewater management, and tracking resource use by accommodation operators. These options remain \emph{qualitative}: the available study does not provide their lifecycle inventories, reduction factors, or modeled uptake. Nor is thermochemical waste treatment automatically emission-free; technology-specific emissions and material balances would be required to establish its environmental benefit.

\subsection{Why Transport Is the Quantified Component}
Road transport was selected because its direct fuel use and corresponding CO$_2$ emissions can be estimated from a small set of engineering inputs. This narrower scope supports an interpretable comparison of ICE and BEV fleets. Nevertheless, the resulting emissions inventory is a \emph{transport operational-use inventory}, not a complete greenhouse-gas inventory for the site. Embedded manufacturing emissions, battery degradation, construction emissions, upstream fuel supply, and non-CO$_2$ gases are excluded from the numerical results.

\section{Study Area, Mobility Technologies, and Input Data}
\subsection{Case-Study Context}
\begin{figure}[!t]
\centering\includegraphics[width=\columnwidth]{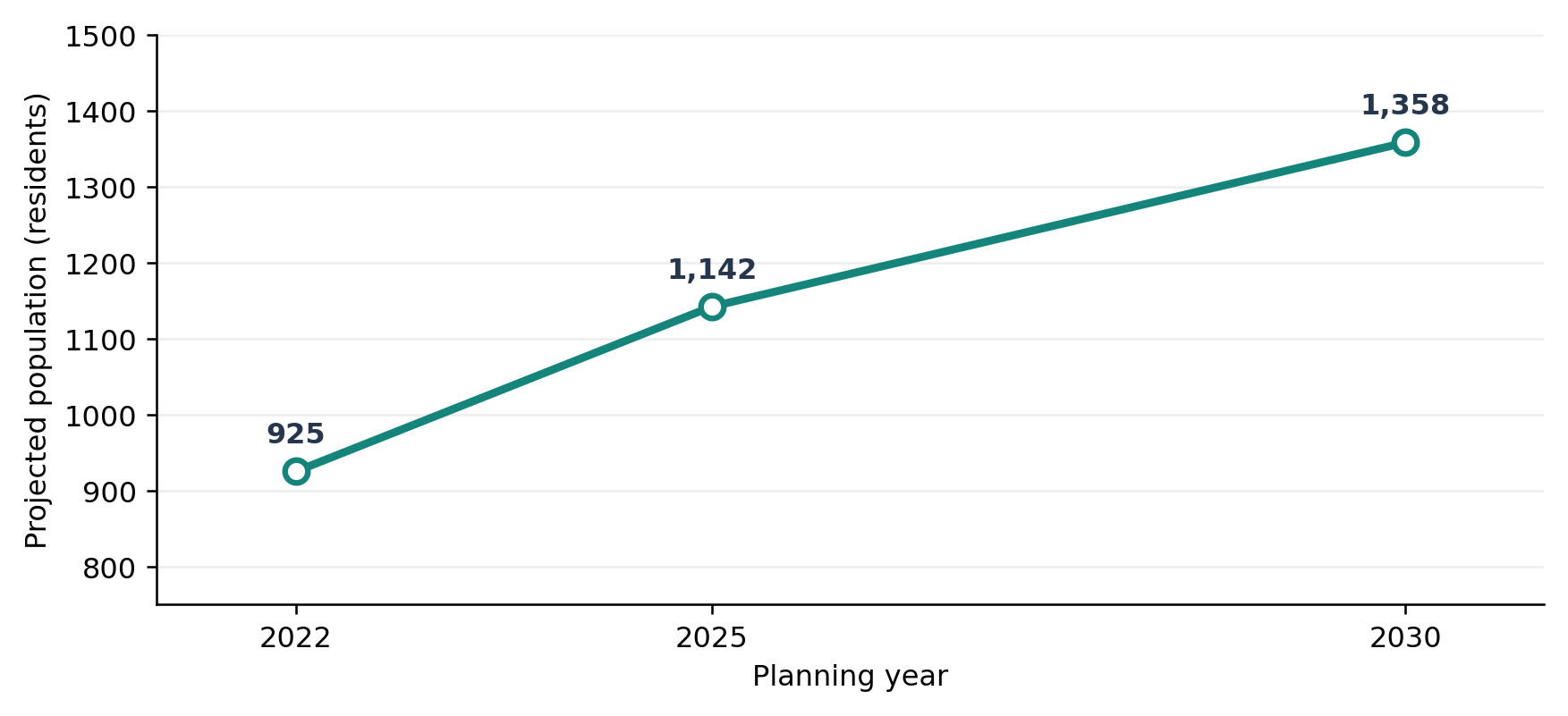}
\caption{Source-study population projections for the planned residential area; 2025 and 2030 are scenario assumptions, not subsequently validated observations.}\label{fig:pop}
\end{figure}

The source design concerns the Hung Phat high-end residential urban development in Duong To, Phu Quoc, within an island-city planning context. The historical study projected the development population from 925 in 2022 to 1,142 in 2025 and 1,358 in 2030 (Fig.~\ref{fig:pop}). These figures are \emph{planning projections originating in the 2022 source}, not observed population counts for those later years. The development road system was described as having an aggregate length of approximately 2.42~km. The original drawings also identified three candidate parking areas that could support electric-vehicle charging; they were conceptual urban-design visualizations rather than a detailed electrical sizing study.

\subsection{Technology Choice}
The original work distinguishes hybrid electric vehicles (HEVs), plug-in hybrid electric vehicles (PHEVs), range-extended electric vehicles (REEVs), battery-electric vehicles (BEVs), and fuel-cell electric vehicles (FCEVs). HEVs and PHEVs retain an ICE, while an REEV may use an engine-driven generator to extend the driving range. An FCEV uses hydrogen or another fuel-cell feedstock; a BEV relies on externally supplied electricity stored in a rechargeable battery. A plug-in hybrid is not equivalent to a non-plug-in HEV, and these categories must not be conflated.

BEVs are the quantitative focus because their use-phase energy input can be represented directly in kWh/km. They also create infrastructure requirements: accessible charging locations, adequate distribution-grid capacity, managed charging to mitigate coincident demand, and appropriate metering. Optional vehicle-to-grid operation could offer flexibility, but its feasibility depends on compatible bidirectional hardware, grid interconnection rules, battery warranty conditions, and user participation. No charging-station siting, grid power-flow, cost, or vehicle-to-grid optimization is performed in the source calculation.

\subsection{Fleet and Mobility Assumptions}
The original projection uses approximately 610 motorcycles and 46 private cars per 1,000 residents, and 15 minibuses per 1,000 residents. The development-level motorcycle mix is assumed to be 60\% scooters and 40\% manual-transmission motorcycles; four- and seven-seat passenger cars are allocated equally. Taxi figures were derived separately from island-wide growth assumptions reported in the source. Such activity coefficients must be understood as modeled assumptions, because the original fleet-development tables and full source statistics are not included in the 21-page document supplied for revision.

The adopted annual motorcycle travel distance is 800~km per vehicle, and the adopted private-car distance is 1,300~km per vehicle, both reflecting movement allocated to the planned area rather than full annual vehicle mileage. Taxi activity is represented by 5~km/day per vehicle. Minibus operation is described as 7~km/day per vehicle with an additional indication of 10 outward and 10 return trips, creating uncertainty about whether 7~km denotes vehicle total distance, route length, or distance per trip. The present analysis retains the source emissions totals instead of imposing an unverified interpretation of these operational details.

\begin{table}[!t]
\caption{Planning Assumptions Explicitly Reported in the Source}
\label{tab:assumptions}\centering\footnotesize
\begin{tabular}{@{}p{.52\columnwidth}p{.38\columnwidth}@{}}\toprule
Item & Source assumption\\\midrule
Dwellings; 2022 residents & 185; 925\\
Population, 2025; 2030 & 1,142; 1,358\\
Motorcycles per 1,000 residents & 610\\
Private cars per 1,000 residents & 46\\
Minibuses per 1,000 residents & 15\\
Scooters : manual motorcycles & 60 : 40\\
Four-seat : seven-seat cars & 50 : 50\\
Motorcycle activity & 800 km/year/vehicle\\
Private-car activity & 1,300 km/year/vehicle\\
Taxi activity & 5 km/day/vehicle\\
Minibus activity & ambiguous; see text\\
Nominal BEV substitutions & 25\%, 50\%, 100\%\\\bottomrule
\end{tabular}
\end{table}

\section{Operational CO$_2$ Accounting Model}
\subsection{Baseline ICE Emissions}
For a vehicle class $i$, the annual use-phase CO$_2$ emissions from combustion may be estimated as
\begin{equation}\label{eq:ice}
E^{\mathrm{ICE}}_{i,y}=\frac{N_{i,y}d_{i,y}f_i\gamma_{s(i)}}{1000},
\end{equation}
where $N_{i,y}$ is the number of class-$i$ vehicles in year $y$, $d_{i,y}$ is the assigned annual travel distance (km/vehicle/year), $f_i$ denotes liquid-fuel consumption (L/km), $\gamma_{s(i)}$ is the emission coefficient for fuel type $s$ (kgCO$_2$/L), and the factor 1000 converts kg to tCO$_2$. This is the fleet-aggregated, unit-consistent form of the source formula $A\times B\times C$ \cite{ipcc2006}.

The fuel-specific CO$_2$ coefficient is constructed from fuel density, net calorific value, and the standard combustion emission factor:
\begin{equation}\label{eq:factor}
\gamma_s=\frac{\rho_s\,\mathrm{NCV}_s\,\mathrm{EF}_s}{1000},
\end{equation}
where $\rho_s$ is in kg/L, $\mathrm{NCV}_s$ in GJ/t, and $\mathrm{EF}_s$ in tCO$_2$/TJ. Equation~\eqref{eq:factor} reproduces the numerical conversion convention of the original work and is consistent with the energy-based mobile-combustion inventory approach \cite{ipcc2006}. The source factors are listed in Table~\ref{tab:fuel}; they reflect a specific assumed fuel composition, and actual fuel-specific factors may differ.

\begin{table}[!t]\caption{Fuel Properties and CO$_2$ Coefficients}\label{tab:fuel}\centering\footnotesize
\begin{tabular}{@{}lrrrr@{}}\toprule
Fuel & $\rho$ & NCV & EF & $\gamma$\\
 & kg/L & GJ/t & tCO$_2$/TJ & kgCO$_2$/L\\\midrule
RON95 gasoline & 0.7407 & 44.75 & 69.30 & 2.30\\
Diesel & 0.8439 & 43.38 & 74.10 & 2.71\\\bottomrule
\end{tabular}
\end{table}

The aggregate ICE reference emissions in year $y$ are
\begin{equation}\label{eq:base}
E^{\mathrm{base}}_y=\sum_{i\in\mathcal V}E^{\mathrm{ICE}}_{i,y},
\end{equation}
where $\mathcal V$ denotes the motorcycle, minibus, passenger-car, and taxi classes. The results in Table~\ref{tab:baseline} are the original study's published class totals. The detailed fuel economy values used for each class were contained in appendices cited by the original text but are absent from the supplied PDF; as a consequence, Eq.~\eqref{eq:ice} can be audited dimensionally but the baseline cannot be independently reconstructed from the available primary input records.

\begin{table}[!t]\caption{Source-Reported Baseline Road-Transport Emissions (tCO$_2$/year)}\label{tab:baseline}\centering\footnotesize
\begin{tabular}{@{}lrrr@{}}\toprule
Vehicle class & 2022 & 2025 & 2030\\\midrule
Manual motorcycle & 9.15 & 11.29 & 13.40\\
Scooter & 18.65 & 23.10 & 27.43\\
Minibus (diesel) & 17.44 & 21.19 & 24.93\\
Four-seat car (gasoline) & 5.01 & 6.45 & 7.41\\
Seven-seat car (gasoline) & 8.16 & 10.49 & 12.05\\
Taxi (gasoline) & 1.00 & 1.34 & 2.01\\\midrule
\textbf{Total} & \textbf{59.41} & \textbf{73.86} & \textbf{87.23}\\\bottomrule
\end{tabular}
\end{table}

\subsection{BEV Electricity-Related Emissions}
For the same class and service demand, the electric-driving counterpart is
\begin{equation}\label{eq:ev}
E^{\mathrm{EV}}_{i,y}=\frac{N_{i,y}d_{i,y}e_{i,y}g_y}{1000},
\end{equation}
where $e_{i,y}$ is electricity consumption (kWh/km) and $g_y$ is the electricity-grid emission factor (kgCO$_2$/kWh). The model counts emissions associated with the electricity used, \emph{not} EV tailpipe emissions. Charging losses must be treated consistently: if $e_{i,y}$ is measured at the battery rather than the charger input, those losses require an additional adjustment. The available source text references, but does not reproduce, the per-class EV energy-use table and all year-specific grid factors.

For context, Vietnam's officially cited 2020 national-grid factor was approximately 0.8041 tCO$_2$/MWh, numerically equivalent to 0.8041 kgCO$_2$/kWh \cite{vietnamfactor}. This historical factor illustrates the relevant order of magnitude; it must \emph{not} be treated as independent verification that the original calculation applied one identical factor in all three study years.

\subsection{Electrification Scenarios and Avoided Emissions}
Let $p_{i,y}$ denote the fraction of class $i$ served by BEVs in year $y$. The mixed-fleet operational emissions are
\begin{equation}\label{eq:mixed}
E^{\mathrm{mix}}_y=\sum_{i\in\mathcal V}\left[(1-p_{i,y})E^{\mathrm{ICE}}_{i,y}+p_{i,y}E^{\mathrm{EV}}_{i,y}\right],
\end{equation}
where the class-level emissions in Eqs.~\eqref{eq:ice} and \eqref{eq:ev} are defined for the same all-ICE or all-BEV fleet size. Avoided operational emissions and the percentage reduction are
\begin{align}
\Delta E_y &= E^{\mathrm{base}}_y-E^{\mathrm{mix}}_y,\label{eq:saving}\\
R_y &=100\Delta E_y/E^{\mathrm{base}}_y.\label{eq:percent}
\end{align}
The original scenarios assign nominal replacement levels of 25\% to 2022, 50\% to 2025, and 100\% to 2030. As the substitution level and planning year increase simultaneously, the reported differences \emph{cannot} isolate the causal impact of BEV penetration from population growth, vehicle mix, fleet activity, or changes in electricity intensity. A controlled comparison would hold the year fixed while varying BEV penetration.

\begin{figure}[!t]
\centering\includegraphics[width=\columnwidth]{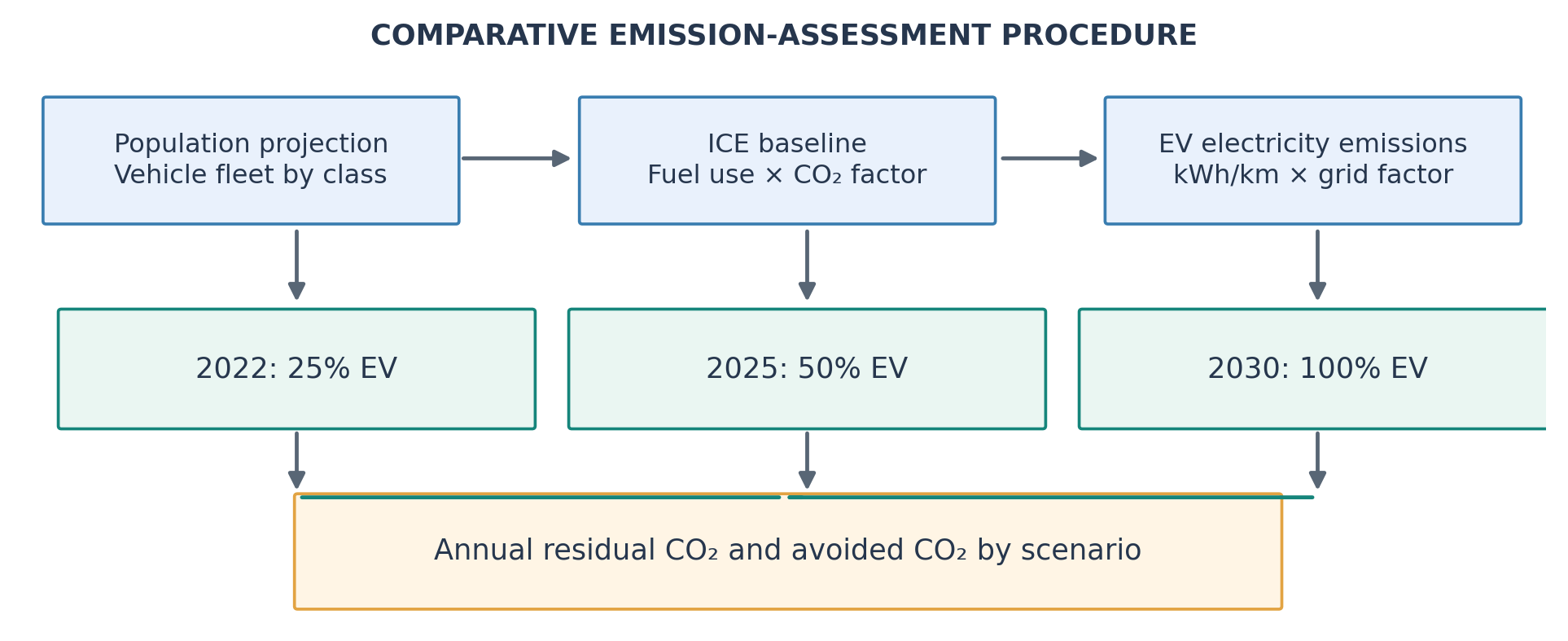}
\caption{Assessment workflow reconstructed in English. Separate calculations for ICE and BEV use the same class-level travel demand in each scenario.}\label{fig:workflow}
\end{figure}

Figure~\ref{fig:workflow} translates the source's nine-stage calculation flow into a compact engineering workflow. Fleet projection feeds the fuel-based reference case and electricity-based EV case, after which each replacement scenario is compared with its own same-year reference.

\section{Scenario Results}
\subsection{Baseline Transport Inventory}
\begin{figure}[!t]
\centering\includegraphics[width=\columnwidth]{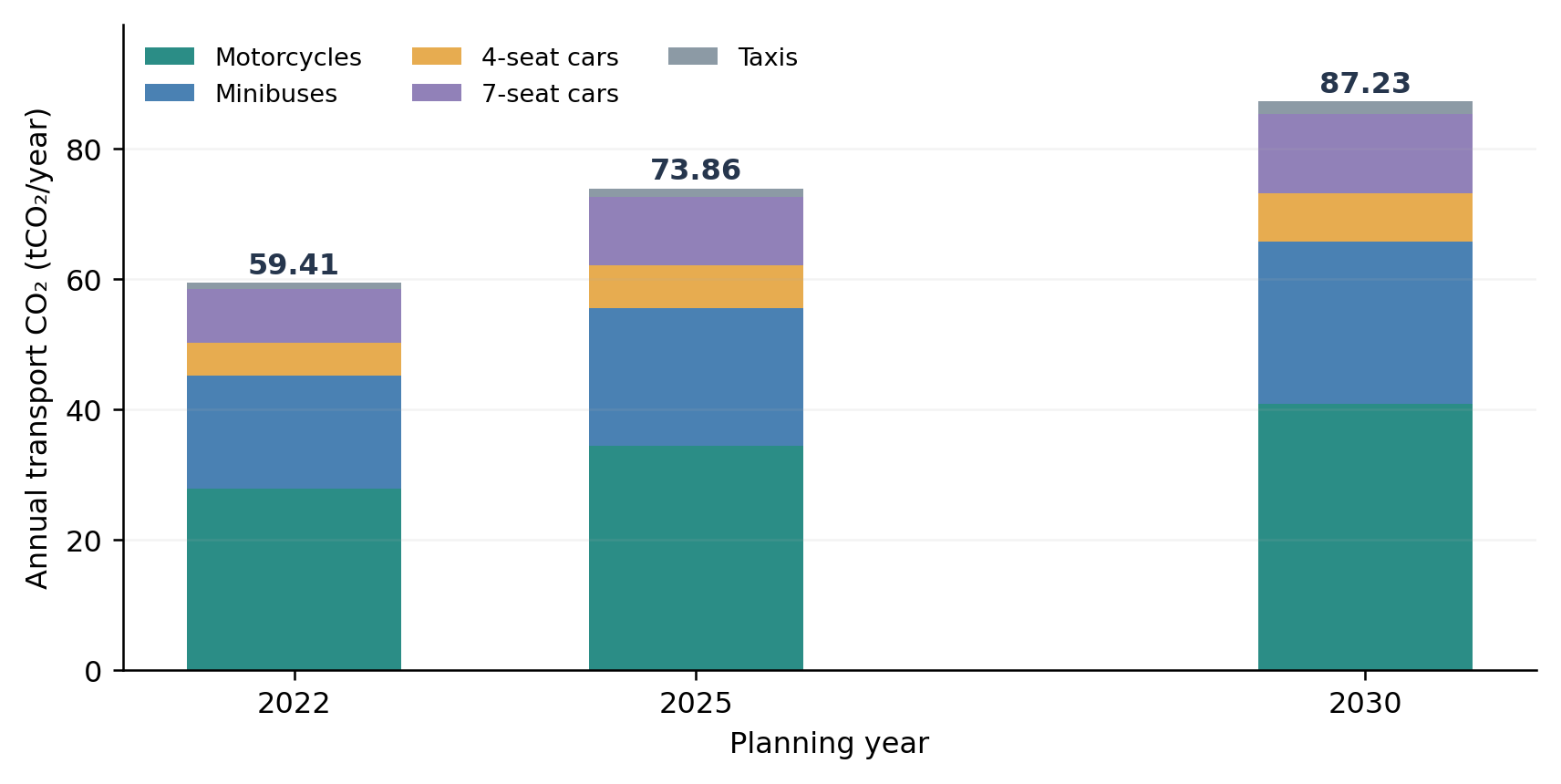}
\caption{Projected reference-case operational road-transport CO$_2$ emissions by vehicle class (source study, 2022).}\label{fig:base}
\end{figure}

The source-reported baseline rises from 59.41 tCO$_2$/year in 2022 to 87.23 tCO$_2$/year in 2030, an absolute increase of 27.82 tCO$_2$/year (46.8\%). Motorcycles form the largest component, followed by minibuses (Fig.~\ref{fig:base}). In 2030 the projected motorcycles contribute 40.83 tCO$_2$/year, and minibuses contribute 24.93 tCO$_2$/year. Their combined contribution is 65.76 tCO$_2$/year, or approximately 75.4\% of the baseline total. This concentration suggests that the aggregate benefit depends strongly on the feasibility of electrifying two-wheelers and high-usage shared-transport vehicles, although these figures are not independent observations of present-day traffic.

\subsection{Vehicle Substitution and Residual Emissions}
Table~\ref{tab:fleet} retains the source-reported BEV vehicle counts for the three scenarios. The 2022 scenario substitutes 141 motorcycles, four minibuses, six four-seat cars, six seven-seat cars, and one taxi. The corresponding 2025 counts are 349, nine, 13, 14, and two vehicles; by 2030 they reach 828, 20, 31, 31, and six vehicles, respectively. These values are forecast or scenario assignments, not recorded fleet registration statistics.

\begin{table}[!t]\caption{BEV Counts Assigned to the Three Source Scenarios}\label{tab:fleet}\centering\footnotesize
\begin{tabular}{@{}lrrr@{}}\toprule
Vehicle class & 2022, 25\% & 2025, 50\% & 2030, 100\%\\\midrule
Motorcycles (both types) & 141 & 349 & 828\\
Minibuses & 4 & 9 & 20\\
Four-seat cars & 6 & 13 & 31\\
Seven-seat cars & 6 & 14 & 31\\
Taxis & 1 & 2 & 6\\\bottomrule
\end{tabular}
\end{table}

\begin{figure}[!t]
\centering\includegraphics[width=\columnwidth]{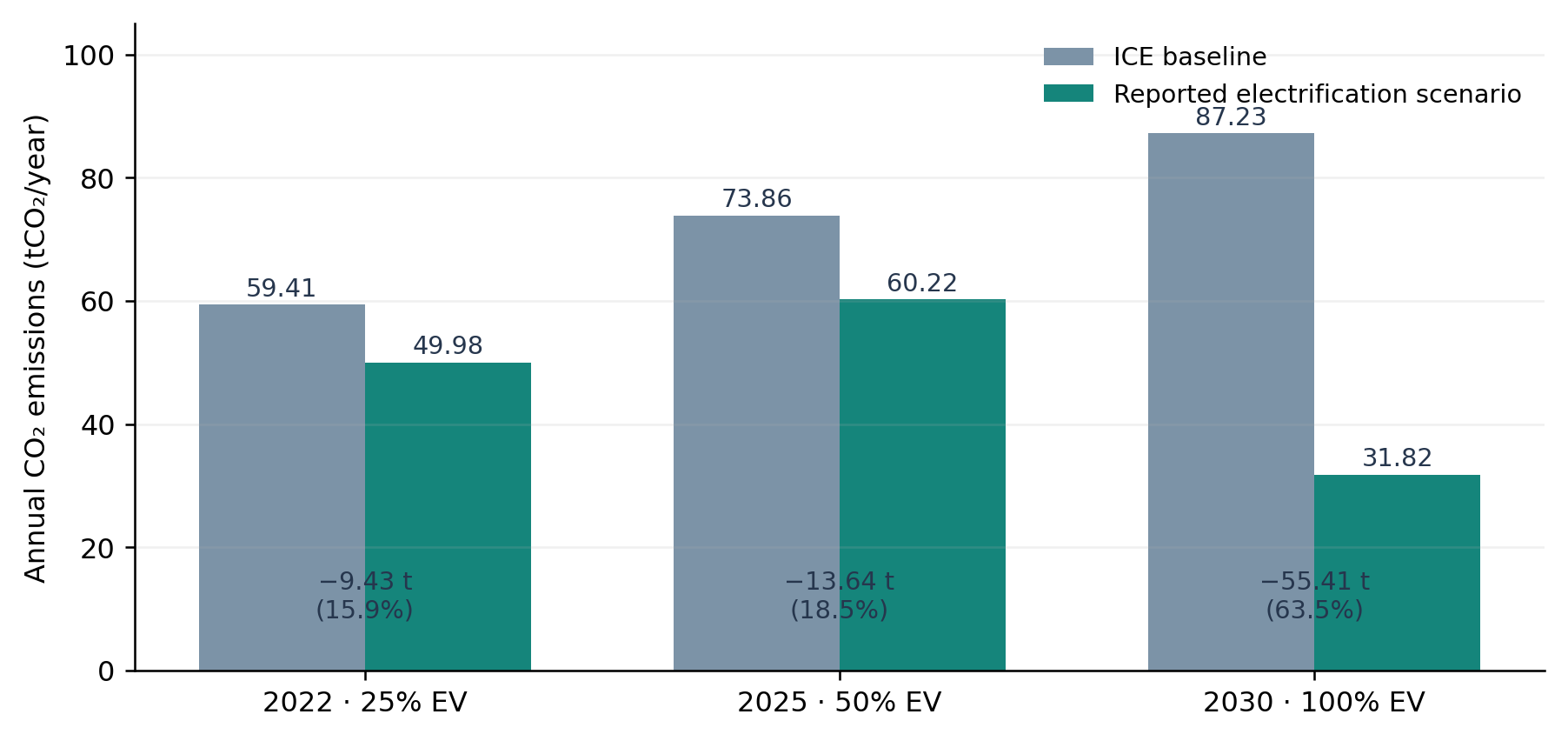}
\caption{Reference and source-reported scenario emissions. Each replacement share applies to a \emph{different year}; the bars are not a same-year sensitivity analysis.}\label{fig:scenario}
\end{figure}

Table~\ref{tab:scenarios} consolidates the original per-class emissions from the three replacement cases. At 25\% substitution in 2022, the source estimates residual emissions of 49.98 tCO$_2$/year, a reduction of 9.43 tCO$_2$/year. At 50\% substitution in 2025, the source reports 60.22 tCO$_2$/year and a reduction of 13.64 tCO$_2$/year. At complete substitution in 2030, it reports 31.82 tCO$_2$/year, a reduction of 55.41 tCO$_2$/year. Figure~\ref{fig:scenario} contrasts these values with their respective baselines. The 2030 BEV-only scenario still generates substantial electricity-related emissions; it is not a zero-CO$_2$ operational scenario on the assumed electricity supply.

\begin{table*}[!t]\caption{Source-Reported Mixed-Fleet Emissions by Vehicle Class (tCO$_2$/year)}\label{tab:scenarios}\centering\footnotesize
\begin{tabular}{@{}lrrr rrr rrr@{}}\toprule
 & \multicolumn{3}{c}{2022: 25\% BEV} & \multicolumn{3}{c}{2025: 50\% BEV} & \multicolumn{3}{c}{2030: 100\% BEV}\\
\cmidrule(lr){2-4}\cmidrule(lr){5-7}\cmidrule(l){8-10}
Class & ICE base & Mixed & Avoided & ICE base & Mixed & Avoided & ICE base & BEV & Avoided\\\midrule
Motorcycles & 27.80 & 22.60 & 5.20 & 34.39 & 27.93 & 6.46 & 40.83 & 10.24 & 30.59\\
Minibuses & 17.44 & 14.62 & 2.82 & 21.19 & 17.75 & 3.44 & 24.93 & 8.75 & 16.18\\
Four-seat cars & 5.01 & 4.58 & 0.43 & 6.45 & 5.67 & 0.78 & 7.41 & 4.83 & 2.58\\
Seven-seat cars & 8.16 & 7.26 & 0.90 & 10.49 & 7.87 & 2.62 & 12.05 & 6.69 & 5.36\\
Taxis & 1.00 & 0.92 & 0.08 & 1.34 & 1.00 & 0.34 & 2.01 & 1.31 & 0.70\\\midrule
\textbf{Total} & \textbf{59.41} & \textbf{49.98} & \textbf{9.43} & \textbf{73.86} & \textbf{60.22} & \textbf{13.64} & \textbf{87.23} & \textbf{31.82} & \textbf{55.41}\\\bottomrule
\end{tabular}
\end{table*}

\subsection{Cross-Scenario Consistency Check}
An important limitation emerges when the 2025 results are checked against the same basic model used for 2022 and 2030. Define the class-specific BEV-to-ICE emission-intensity ratio
\begin{equation}\label{eq:q}
q_i=\frac{E^{\mathrm{EV}}_{i,2030}}{E^{\mathrm{ICE}}_{i,2030}},
\end{equation}
where the numerator is the reported 2030 complete-BEV class emission and the denominator is that year's all-ICE class emission. Values inferred from Table~\ref{tab:scenarios} are approximately 0.251 for motorcycles, 0.351 for minibuses, 0.652 for four-seat cars, 0.555 for seven-seat cars, and 0.652 for taxis. These are \emph{derived diagnostic ratios}, not independently measured efficiencies.

If a common ratio $q_i$ is assumed across the scenarios and the same substitution fraction applies within each vehicle class, a purely diagnostic counterfactual is
\begin{equation}\label{eq:diag}
\widehat{E}^{\mathrm{mix}}_y=\sum_i E^{\mathrm{ICE}}_{i,y}\bigl[(1-p_y)+p_yq_i\bigr].
\end{equation}
Using $p_{2025}=0.5$ gives $\widehat{E}^{\mathrm{mix}}_{2025}=50.41$ tCO$_2$/year and avoided emissions of 23.45 tCO$_2$/year, rather than the source-reported 60.22 and 13.64 tCO$_2$/year. For 2022, this diagnostic gives 49.94 tCO$_2$/year, close to the reported 49.98, with the difference consistent with rounding and small class differences. In 2030 the model reproduces the source data by construction. Table~\ref{tab:audit} makes this distinction explicit.

\begin{table}[!t]\caption{Internal Consistency Diagnostic; Not a Replacement for Source Results}\label{tab:audit}\centering\footnotesize
\begin{tabular}{@{}lrrrr@{}}\toprule
Year & Base & Reported mix & Diagnostic mix & Difference\\\midrule
2022 & 59.41 & 49.98 & 49.94 & $-0.04$\\
2025 & 73.86 & 60.22 & 50.41 & $-9.81$\\
2030 & 87.23 & 31.82 & 31.82 & 0.00\\\bottomrule
\end{tabular}
\\[2pt]\parbox{.94\columnwidth}{\footnotesize \emph{Note:} Diagnostic values assume that the per-class BEV/ICE emission-intensity ratios estimated from the 2030 all-BEV case also apply in 2022 and 2025, with uniform penetration within each class. This assumption is not established by the source; hence the diagnostic does not constitute a corrected dataset.}
\end{table}

The discrepancy could reflect a changed grid-emission factor, different annual travel assumptions, a class-allocation choice, or an arithmetic/data-transfer error. The available document does not contain the detailed EV energy-consumption and grid-factor appendix needed to discriminate among these explanations. It would be scientifically unjustified either to silently ``correct'' the 2025 source value or to treat it as independently validated. Therefore, the scenario reduction figures are retained as \emph{source-reported projections subject to verification}.

\subsection{Break-Even Grid Intensity and Analytical Sensitivity}
The accounting model also provides a simple test for whether substitution lowers operational CO$_2$ emissions in a particular vehicle class. Equations~\eqref{eq:ice} and \eqref{eq:ev} imply that BEV operation has lower use-phase CO$_2$ emissions than an equivalent ICE vehicle whenever
\begin{equation}\label{eq:breakeven}
g_y < g^{\star}_{i,y}=\frac{f_i\gamma_{s(i)}}{e_{i,y}},
\end{equation}
where $g^{\star}_{i,y}$ is the class-specific break-even grid-emission factor. This threshold depends on directly comparable service demand, consistent charging-loss treatment, and actual energy intensity. It does not establish lifecycle superiority, which additionally depends on manufacturing and upstream supply chains.

For any fixed fleet and substitution fractions, the avoided-emissions sensitivity to the assumed grid-emission factor is
\begin{equation}\label{eq:gridsens}
\frac{\partial\Delta E_y}{\partial g_y}=-\frac{1}{1000}\sum_{i\in\mathcal V}p_{i,y}N_{i,y}d_{i,y}e_{i,y}.
\end{equation}
Thus, within the proposed operational accounting boundary, the avoided CO$_2$ quantity decreases linearly as the grid factor increases. However, numerical break-even factors or sensitivity bands cannot be responsibly reported without the missing class-level EV consumption parameters. They are derived relationships, not additional case-study results.

The same distinction matters at the urban scale. A citywide net balance would require an accounting identity of the form
\begin{equation}\label{eq:citywide}
E_y^{\mathrm{net}}=E_y^{\mathrm{transport}}+E_y^{\mathrm{buildings}}+E_y^{\mathrm{construction}}+E_y^{\mathrm{waste}}+E_y^{\mathrm{other}}-R_y,
\end{equation}
where consistent system boundaries, emission scopes, and verified anthropogenic removals $R_y$ are essential. Only the first component is evaluated numerically in this paper; therefore Eq.~\eqref{eq:citywide} is a scope reminder, not a computed net-zero balance.

\section{Planning Interpretation and Infrastructure Implications}
\subsection{Electricity Decarbonization Matters}
The 2030 all-BEV projection leaves 31.82 tCO$_2$/year in electricity-related emissions. This residual highlights the importance of electricity emissions rather than tailpipe emissions alone. Low-carbon electricity procurement, local renewable generation, and time-aware charging can potentially reduce the operational carbon intensity of charging, although their actual benefit depends on generation mix, system dispatch, and boundary conventions. A claim of ``zero-emission charging'' would require an explicit, defensible temporal and accounting basis, not merely the presence of a rooftop photovoltaic array.

\subsection{Charging Infrastructure as a Separate Planning Problem}
\begin{figure}[!t]
\centering\includegraphics[width=\columnwidth]{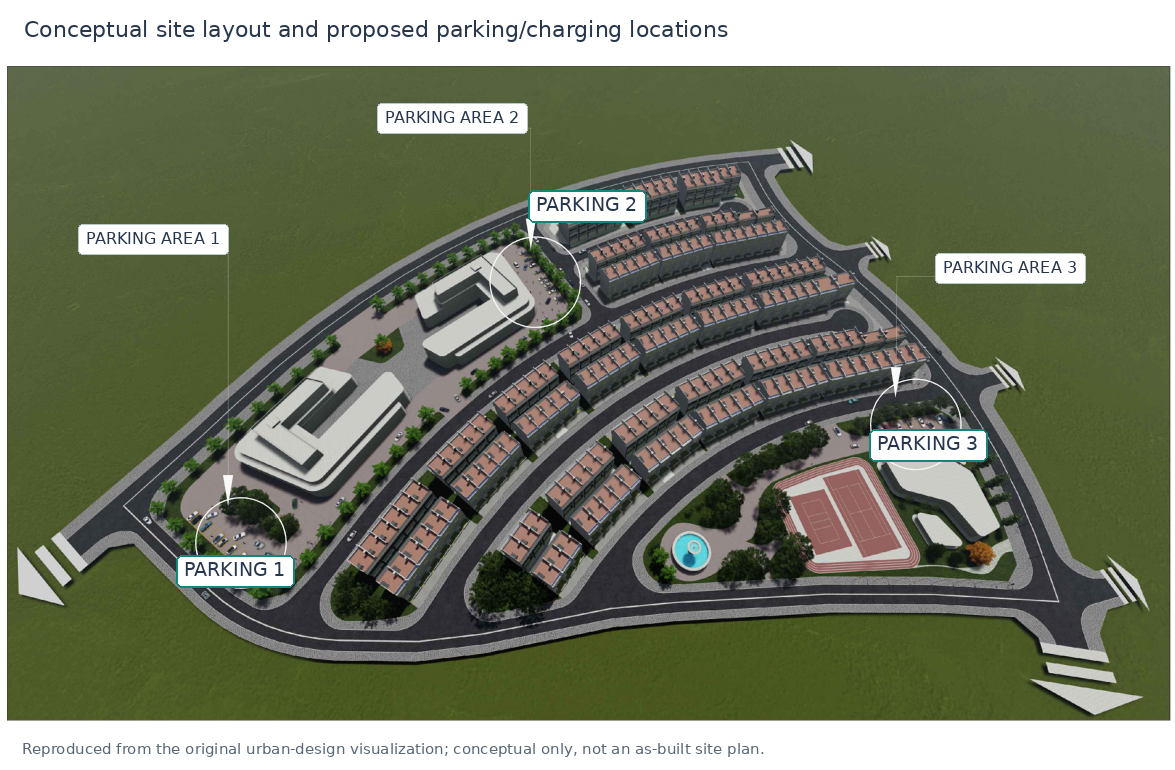}
\caption{English-annotated conceptual layout reproduced from the original urban-design visualization, showing three proposed parking areas. These annotations are illustrative; original renderings do not establish electrical design approval.}\label{fig:site}
\end{figure}

The source's visualization identifies three proposed parking locations that could potentially host charging facilities (Fig.~\ref{fig:site}). This is a useful architectural starting point but is insufficient to determine charger number, AC versus DC technology, rated power, accessibility, tariff design, or electrical connection feasibility. Those decisions require an EV arrival and dwell-time model, demand diversity, charging-duration requirements, vehicle fleet composition, feeder and transformer capacities, voltage and thermal limits, and reliability assessments. A subsequent infrastructure study should quantify coincident charging demand and test its impact on Phu Quoc's electricity supply and distribution network rather than equating vehicle replacement counts with charging-point counts.

\subsection{Complementary Urban Interventions}
Walking and cycling infrastructure, appropriately designed public transit, and reduced travel demand may lower emissions even before full fleet replacement. Building-energy efficiency, recycled materials, efficient waste collection, low-carbon water services, and nature conservation act in different accounting categories. A scientifically complete net-zero plan must avoid double counting between electricity savings, local power generation, and externally reported emission reductions. Each action requires an activity baseline, emission factors, implementation constraints, and a method for ongoing measurement and verification.

\subsection{Priorities for a Defensible Follow-On Study}
A next-stage study should: (i) establish a geospatially defined annual urban inventory consistent with the Global Protocol for Community-Scale Greenhouse Gas Emission Inventories; (ii) obtain observed traffic counts and measured or manufacturer-specific energy consumption by class; (iii) reconstruct and independently reproduce the 2025 mixed-fleet table; (iv) use a clearly justified grid-emission trajectory and disclose charging losses; (v) evaluate equal-year BEV penetration levels and uncertainty intervals rather than confounding time and substitution; and (vi) couple charging demand with feeder-constrained siting and sizing models. These extensions would support robust investment decisions, while lifecycle and non-transport inventories would be necessary to evaluate any sitewide net-zero claim.

\section{Limitations and Reproducibility}
The case study is a retrospective presentation of assumptions developed in 2022. The years 2025 and 2030 are modeled projections, not evidence of observed urban outcomes. The original source refers to technical appendices containing fleet counts, fuel-consumption parameters, EV electricity-use assumptions, grid factors, and comprehensive reduction tables; those appendices were not part of the supplied 21-page material. Accordingly, some numerical results can be checked only by summing disclosed table entries, not by regenerating them from complete primary inputs.

The assessment counts operational CO$_2$ rather than all greenhouse gases or full vehicle lifecycles. Travel distances allocated to a small development may not represent total vehicle activity; the minibus route/trip definition is ambiguous; taxi assumptions require clarification; and rounding may affect the displayed totals. The year-dependent emission factors, charging losses, electricity-system constraints, and actual charging behavior are not fully documented. The 2025 cross-scenario discrepancy, described above, is particularly important. All numerical results should be labeled projections, not field measurements or independently verified emissions savings.

The graphics and original site visualization have been reformatted into English with a light background. The engineering site illustration is conceptual and is not a georeferenced or construction-ready charging infrastructure plan. No additional empirical dataset, grid study, cost model, or numerical simulation was introduced during this editorial reconstruction.

\section{Conclusion}
This study reformulates an urban net-zero planning concept as a bounded comparison of internal-combustion and battery-electric road transport in a planned Phu Quoc residential development. The original study reports baseline CO$_2$ emissions of 59.41, 73.86, and 87.23 tCO$_2$/year in 2022, 2025, and 2030, and corresponding avoided emissions of 9.43, 13.64, and 55.41 tCO$_2$/year under nominal 25\%, 50\%, and 100\% BEV substitution. These results illustrate the potential importance of transport electrification, particularly for motorcycles and minibuses, while confirming that BEVs still carry electricity-related carbon emissions under a nonzero grid factor.

The numerical audit finds that the reported 2025 mixed-fleet emissions are not consistent with time-invariant class-specific BEV/ICE intensity ratios inferred from the other scenarios. Since the detailed energy-input appendices are unavailable, that difference remains unresolved. The main supported conclusion is thus conditional: transport electrification can be a significant part of low-carbon urban development, but the magnitude of savings requires validated input data, comparable scenarios, and transparent electricity accounting. Demonstrating net-zero performance for the entire development additionally requires complete sectoral inventories, robust carbon-removal accounting, and an infrastructure-feasible implementation plan.

\section*{Source and Editorial Transparency}
The scenario data, population assumptions, transport parameters, and architectural visualization originate from T.-K. Nguyen's 2022 study \cite{originalstudy}. This English revision restructures and technically reviews those materials. New diagnostic computations are expressly identified and do not replace the primary reported dataset. Other references provide background standards and planning context, not independent confirmation of the case-study scenario values.

\balance

\end{document}